\documentclass[a4paper,11pt]{article}
\usepackage{pos}
\usepackage{multirow}
\usepackage{upgreek}
\usepackage{subcaption}
\usepackage{wrapfig}
\usepackage[font=small,skip=0pt]{caption}
\usepackage[inline]{enumitem}
\let\OLDthebibliography\thebibliography
\renewcommand\thebibliography[1]{
  \OLDthebibliography{#1}
  \setlength{\parskip}{1pt}
  \setlength{\itemsep}{1pt plus 0.3ex}
}

\title{Discrimination of muons for mass composition studies of inclined air showers detected with IceTop}
 \ShortTitle{IceTop: Composition of inclined air showers}

\author{The IceCube Collaboration \\{\normalsize \normalfont(a complete list of authors can be found at the end of the proceedings)}}

\emailAdd{aswathi.balagopalv@icecube.wisc.edu}

\abstract{IceTop, the surface array of IceCube, measures air showers from cosmic rays within the energy range of 1 PeV to a few EeV and a zenith angle range of up to $\approx$ 36$^\circ$. This detector array can also measure air showers arriving at larger zenith angles at energies above 20 PeV. Air showers from lighter primaries arriving at the array will produce fewer muons when compared to heavier cosmic-ray primaries. A discrimination of these muons from the electromagnetic component in the shower can therefore allow a measurement of the primary mass. A study to discriminate muons using Monte-Carlo air showers of energies 20-100 PeV and within the zenith angular range of 45$^\circ$-60$^\circ$ will be presented. The discrimination is done using charge and time-based cuts which allows us to select muon-like signals in each shower. The methodology of this analysis, which aims at categorizing the measured air showers as light or heavy on an event-by-event basis, will be discussed.

\vspace{4mm}
{\bfseries Corresponding authors:}
Aswathi Balagopal V.$^{1*}$\\
{$^{1}$ \itshape Wisconsin IceCube Particle Astrophysics Center, University of Wisconsin, Madison, WI 53703, USA}\\[4mm]
$^*$ Presenter

\FullConference{37$^{\rm{th}}$ International Cosmic Ray Conference (ICRC 2021)\\
		July 12th -- 23rd, 2021\\
		Online -- Berlin, Germany}

}

\begin{document}
\maketitle
\section{Introduction}\label{sec:info}
The determination of the primary mass composition of cosmic rays with varying energies provides added information about the origin of the cosmic rays that contribute to the observed cosmic-ray spectrum and their propagation within the Universe. A common method to determine the mass of the cosmic-ray primary is with the use of the muon content within the air shower. Heavy primaries, like iron, will generate a larger number of muons in the air shower than light primaries, like proton, of the same energy.

\begin{wrapfigure}{r}{0.4\textwidth}
  \vspace{-0.8cm}
  \begin{center}
    \includegraphics[width=0.4\textwidth]{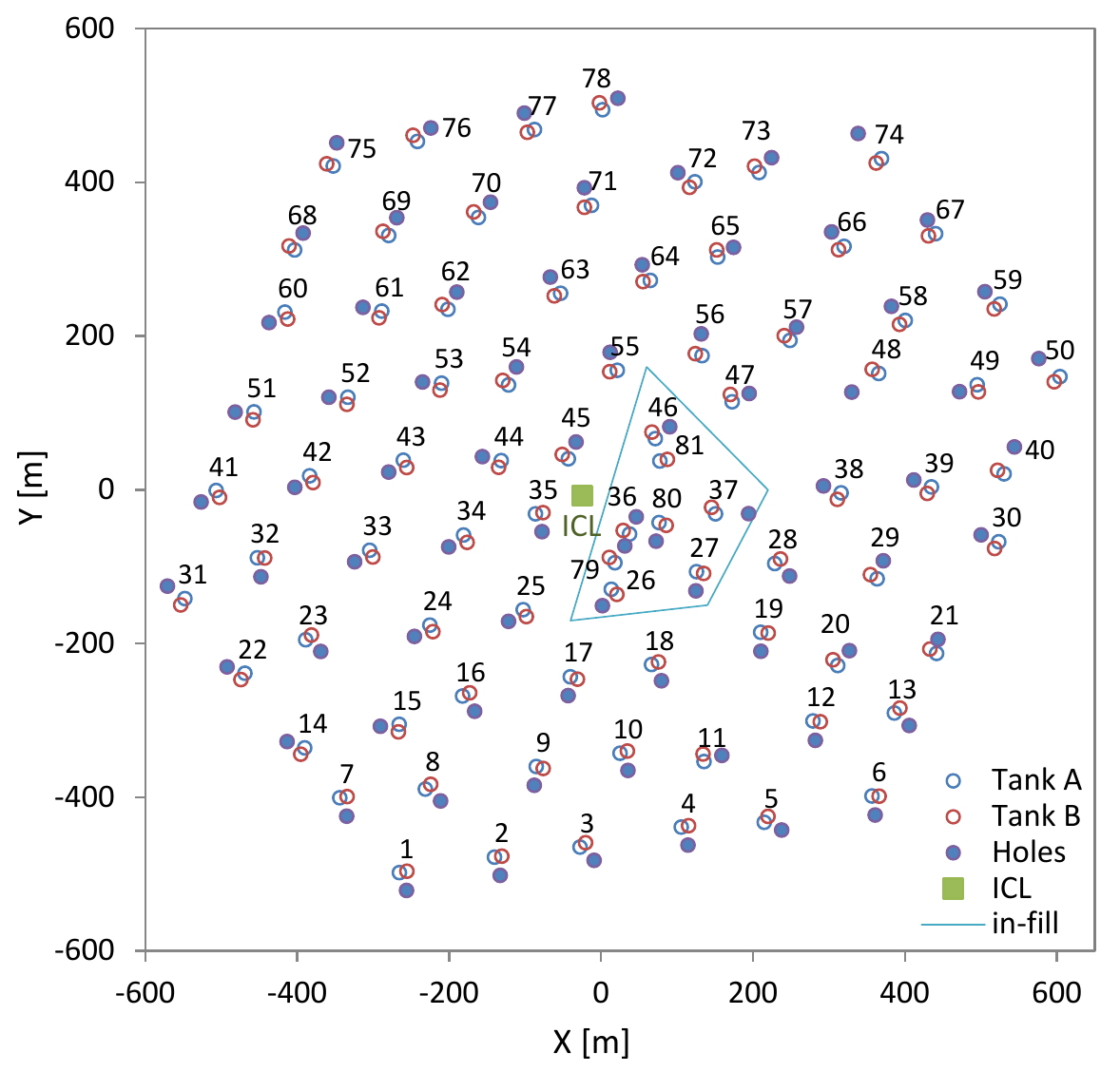}
  \end{center}
  \caption{The layout of the IceTop array from \cite{2013NIMPA.700..188A}.}\label{fig1}
\end{wrapfigure}
In this analysis, we explore a method that can be used to determine the light or heavy mass group of the primary cosmic ray on an event-by-event basis. The study is performed for inclined air showers detected by the IceTop array. Such a study is important for the test of existing hadronic interaction models, and the muon content for different mass groups predicted by such models. The composition study for inclined air showers measured by IceTop will also provide an independent cross-check to other composition analyses of IceTop using quasi-vertical air showers. IceTop is the surface component of IceCube, the neutrino telescope at the South Pole. IceTop consists of 162 ice-Cherenkov tanks that measure signals deposited by charged particles arising from cosmic-ray air showers \cite{2013NIMPA.700..188A}. These 162 tanks are grouped into 81 stations, with each station composed of two tanks. Figure \ref{fig1} shows the geometry and layout of the IceTop array. Each IceTop tank contains two digital optical modules (DOMs) that collect the deposited Cherenkov light. Since Cherenkov light is emitted by both muons and electromagnetic particles (with non-negligible hadronic contribution), a combination of signals from these will be observed in the tank signal. 

There exist several analyses using IceTop and IceCube with the goal of discriminating the electromagnetic particles from the muons. For analyses focused on the determination of the muon density, it is crucial to get a near-perfect number of the observed muons \cite{Soldin:2021icrc}. However, in the cases where the discrimination is used to determine the mass-composition, a contamination of electromagnetic components within the muon parameter can be tolerated \cite{IceCube:2019hmk}.

Other analyses using IceTop and IceCube for mass-composition studies operate only for showers with zenith angles up to $36^\circ$ \cite{Kang:2021icrc, IceCube:2019hmk, Plum:2021icrc, Verpoest:2021icrc}. Air showers with zenith angles ($\theta$) larger than this are generally not included within the standard IceTop analyses. This is primarily due to the unavailability of energy calibration (for the energy proxy S$_{125}$ which is the signal expected to be deposited in a tank at a distance of 125 m) using simulations for these air showers. However, we can determine the energy of these showers with larger $\theta$ using the method of constant intensities, where we compare the inclined showers to the vertical showers (with the expectation of universal flux rates) to determine their energy \cite{BalagopalV2019_1000091377}. Therefore, we can use such inclined air showers also for mass-composition studies; which is done in this analysis.
\section{Analysis method}
We use air-shower simulations that were generated using CORSIKA \cite{Heck:1998vt}, with Sibyll 2.1 \cite{PhysRevD.80.094003} as the hadronic-interaction model. We consider simulations with zenith angles above 45$^\circ$ and energies above 10$^{7.3}$ GeV. This is the energy threshold of inclined air showers measured with IceTop \cite{BalagopalV2019_1000091377}. Since the available simulation set (with full detector response) ends at an energy of 10$^8$ GeV, this is the maximum energy that we consider in this analysis. We divide the showers into three zenith bins: 45$^\circ$-50$^\circ$, 50$^\circ$-55$^\circ$, and 55$^\circ$-60$^\circ$. These showers are also divided into different bins of their energy proxy  log$_{10}$(S$_{125}$/VEM), where VEM is the expected amount of charge deposited in a tank by a vertical-equivalent muon. We choose only the bins of log$_{10}$(S$_{125}$/VEM) where the energy is above 10$^{7.3}$ GeV. Also, some energy-proxy bins are dropped due to insufficient statistics for the simulations. The energy-proxy bins that we use are:
log$_{10}$(S$_{125}$/VEM) = 0.7-1.0, 1.0-1.3, 1.3-1.6 for $45^\circ$-$50^\circ$ showers, and log$_{10}$(S$_{125}$/VEM) = 0.4-0.7, 0.7-1.0, 1.0-1.3 for showers with zenith angles $50^\circ$-$55^\circ$ and $55^\circ$-$60^\circ$ \cite{BalagopalV2019_1000091377}.

We apply cuts to the simulated events to ensure good reconstruction quality for the events used in the analysis. These event-level quality cuts are as follows:
\begin{enumerate*}
    \item We require 5 IceTop stations to be triggered and passed through the filter.
    \item The reconstruction of the air shower is required to have succeeded.
    \item The number of layers of tanks around the shower core should be greater than one. This ensures that at least two outer-layers of tanks are there surrounding the shower core.
    \item The tank with the maximum signal within the shower should not be located at the edge of the array. Even if the previous cut is applied, we sometimes are left with events where the maximum signal is in the edge, due to noisy hits. Application of this cut ensures better quality for the events chosen.
\end{enumerate*}
Although we reject $\approx80\%$ events with these cuts, the remaining events have good quality, which is needed to conduct this study.

Apart from these event-level quality cuts, we apply some analysis-level cuts to select the muon-like signals in the air showers. The first cut is a charge-based cut. We choose those hits in the tanks which have a charge between 1 and 4 VEM. This was chosen since muons from inclined air showers were mainly seen to fall within this charge range for IceTop. This cut already removes a majority of the electromagnetic contamination within the selected hits; especially at larger distances from the shower core. 
\begin{figure}[h]
\includegraphics[width=.5\textwidth]{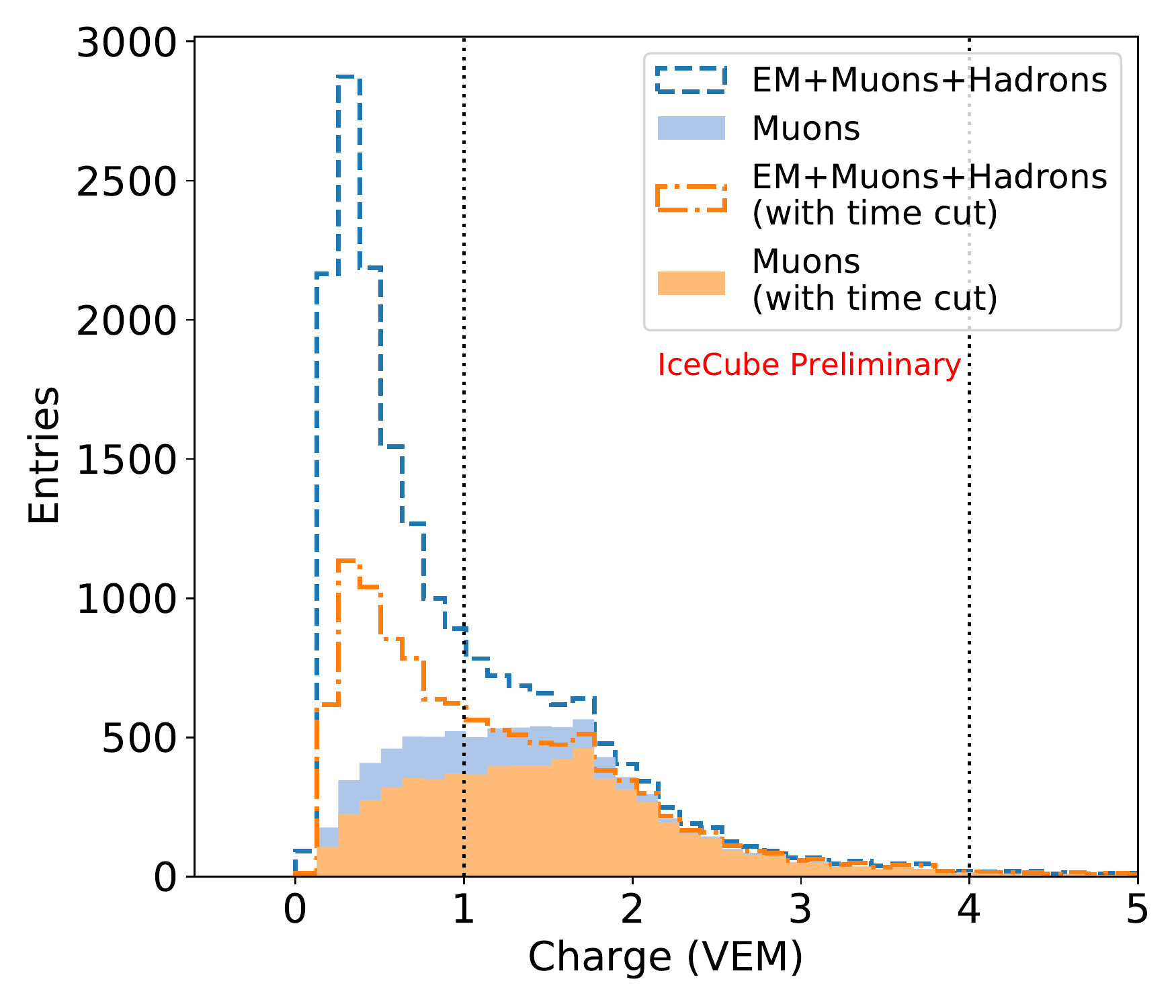}\includegraphics[width=.49\textwidth]{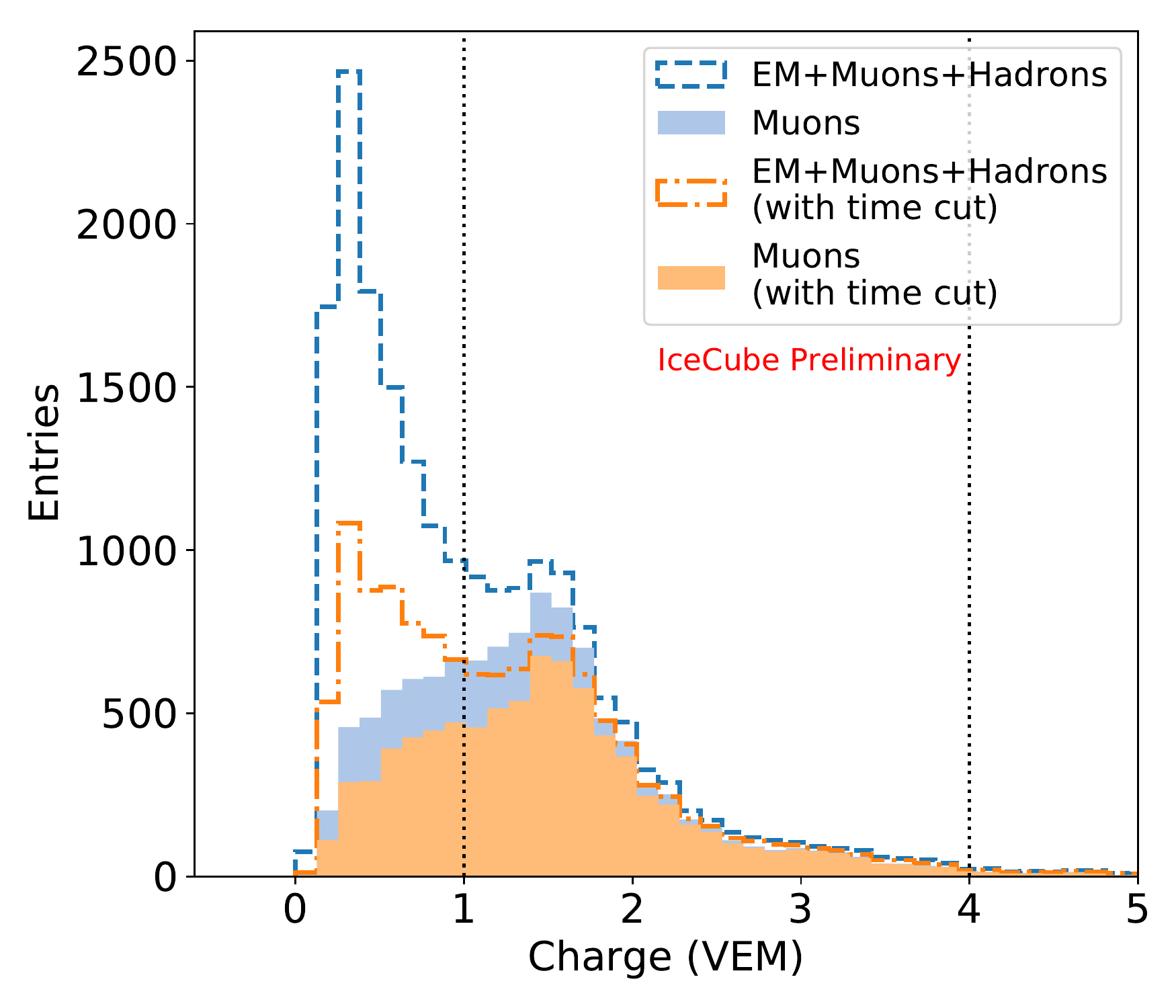}
\caption{Left: Proton showers, $\theta\,=\,50^\circ-55^\circ$ and log$_{10}$(S$_{125}$/VEM)$\,=\,0.7-1.0$. The charge within the tanks from all particle types for hits with distance from the shower axis $\geq 260$ m are shown. These are the events that pass the event-level quality cuts. The muon-component of these hits are also shown. The charge distribution once the time-based cut is applied is also shown (including its muon content). All hits that fall within the two vertical lines between 1 and 4 VEM will be chosen as the final muon-like hits. Right: The same for iron showers, $\theta\,=\,45^\circ-50^\circ$ and log$_{10}$(S$_{125}$/VEM)$\,=\,1.0-1.3$. The hits with distance from the shower axis $\geq 330$ m are shown.}\label{fig2}
\end{figure}
In addition to this, we apply a time-based cut to the remaining hits in the event. Only those events with $\Delta(\delta\mathrm{t}) > 0$ are chosen in the final level. 
Here, $\delta \mathrm{t}$ is the curvature of the shower front. 
$\Delta(\delta\mathrm{t}) = \delta\mathrm{t_{observed}}-\delta\mathrm{t_{predicted}}$ compares the observed curvature to the theoretical prediction of the shower curvature (which is optimised for quasi-vertical showers).
The condition $\Delta(\delta\mathrm{t}) > 0$ 
chooses the hits that arrived before their predicted times. Combined with the charge-based cut, this allows us to choose the early muons in the shower.
The addition of this cut further removes electromagnetic contamination from the selected hits. Figure \ref{fig2} shows the effect of the analysis-level cuts that selects muon-like hits for proton and iron showers.

Once the final-level hits have been selected, they are binned with respect to their perpendicular distance to the shower axis. The sum of all hits above each distance bin is evaluated to obtain the muon-like parameter.
\begin{equation}
    \xi_{\mu-\mathrm{like}}\,=\, \Sigma\, N_{\mathrm{\mu-like}}\,>\,r
\end{equation}

\begin{wrapfigure}{r}{0.6\textwidth}
\vspace{-0.5cm}
\centering
\includegraphics[width=.6\textwidth]{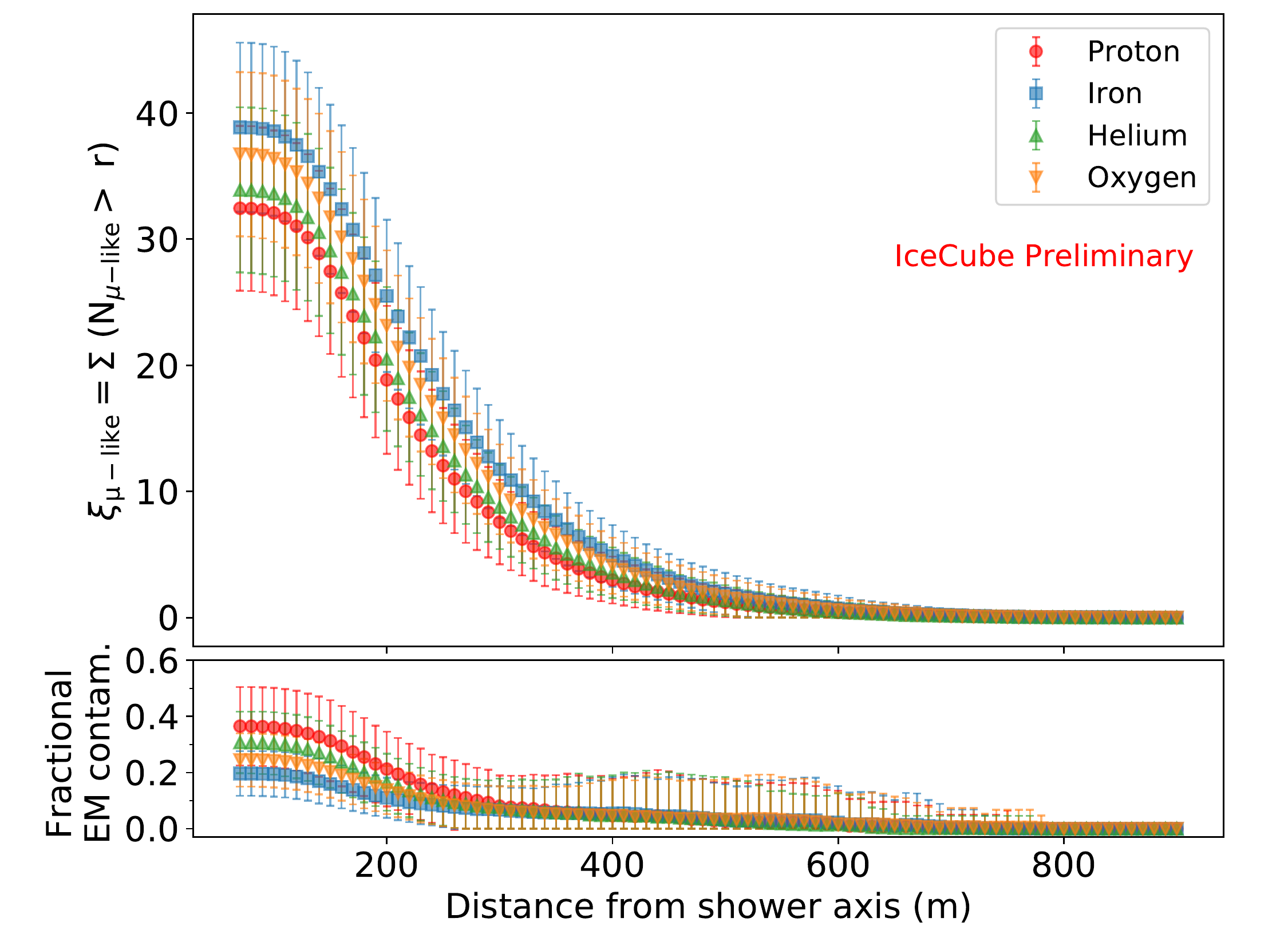}
\caption{Top panel: lateral distribution of the muon-like parameter,  $\xi_{\mu-\mathrm{like}}$, for showers from different primaries with $\theta\,=\,55^\circ-60^\circ$ and log$_{10}$(S$_{125}$/VEM) = 0.7-1.0. Bottom panel: corresponding fraction of electromagnetic contamination within signals considered in the $\xi_{\mu-\mathrm{like}}$ calculation. Both the top and bottom panels show the mean values (marker) and standard deviations (error bar) of the respective distribution.}\label{fig3}
\end{wrapfigure}

The lateral distribution of the muon-like parameter, $ \xi_{\mu-\mathrm{like}}$ for air showers within the zenith angular range of 55$^\circ$-60$^\circ$ for the energy bin log$_{10}$(S$_{125}$/VEM) = 0.7-1.0 is shown in Figure \ref{fig3}. 
This is shown for simulated air showers of different primaries. As expected, the mean of $ \xi_{\mu-\mathrm{like}}$ exhibits the trend where heavier primaries have more muon-like hits than lighter primaries. We can also see that there is a good amount of separation between the distribution of the light and the heavy primaries. We can therefore use this parameter to determine the lightness/heaviness of each shower. The bottom panel of Figure \ref{fig3} depicts the fraction of electromagnetic contamination within $ \xi_{\mu-\mathrm{like}}$ at each distance bin. The electromagnetic contamination is seen to be small, especially at distances further from the center of the shower.

The next step in the analysis is the calculation of the reference distance at which we get the maximum separation between the $ \xi_{\mu-\mathrm{like}}$ of the heavy and light primaries. For this we look at the proton and iron primaries alone. A Gaussian distribution is fit to the $ \xi_{\mu-\mathrm{like}}$ at each distance bin for both the proton and iron showers. A measure of the separation of these two distributions is estimated to get the distance at which the separation is maximum, with least amount of fluctuations. For this, the figure of merit (FOM) is calculated as
\begin{equation}
    \mathrm{FOM}\,=\, \frac{|\mathrm{\upmu(H)-\upmu(Fe)}|}{\sqrt{\mathrm{\sigma_{H}^2+\sigma_{Fe}^2}}},
\end{equation}
where $\upmu$ is the mean of the Gaussian distribution and $\sigma$ is the standard deviation \cite{Holt:2019fnj}.

\begin{figure}[h]

\includegraphics[width=.5\textwidth]{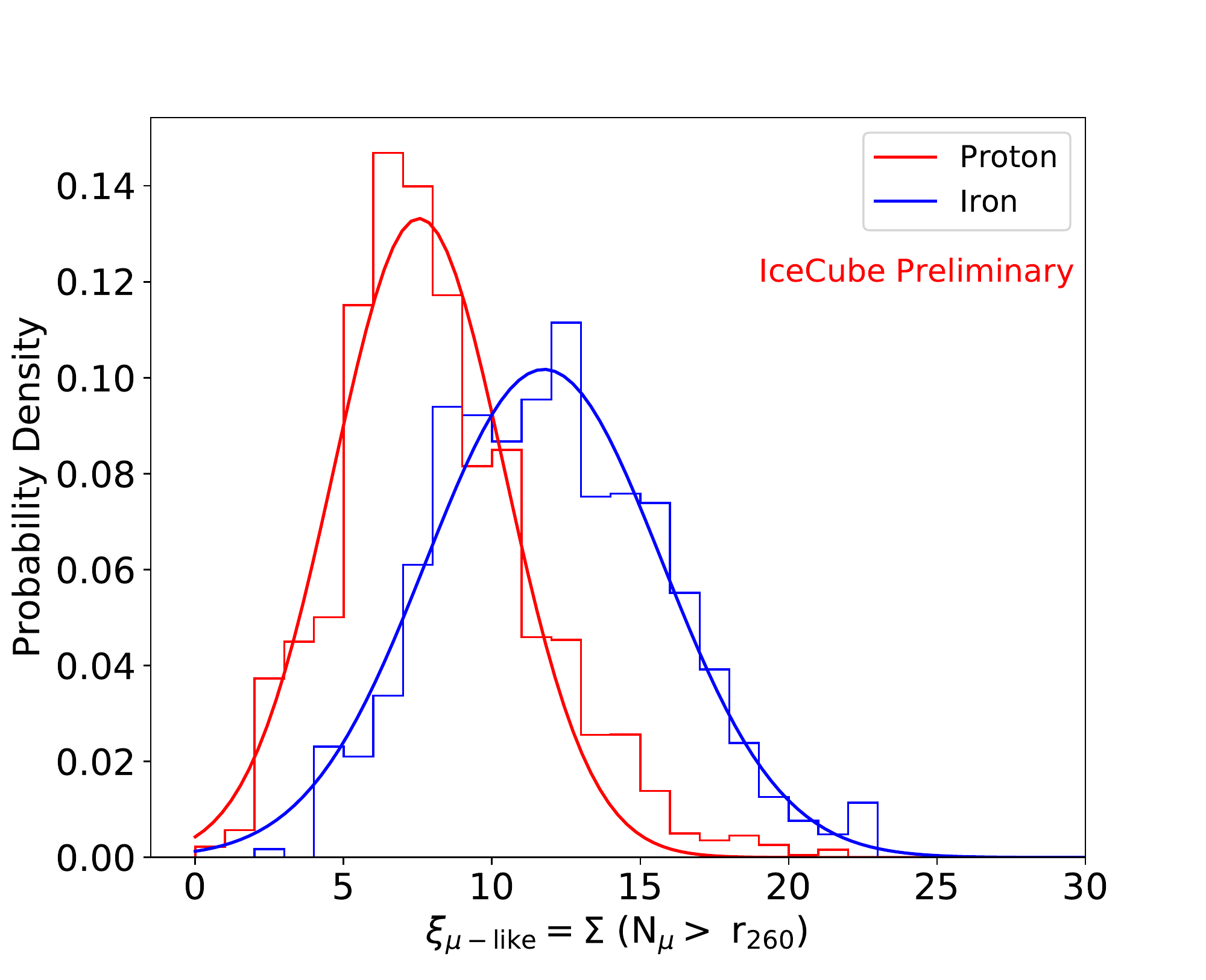}\includegraphics[width=.49\textwidth]{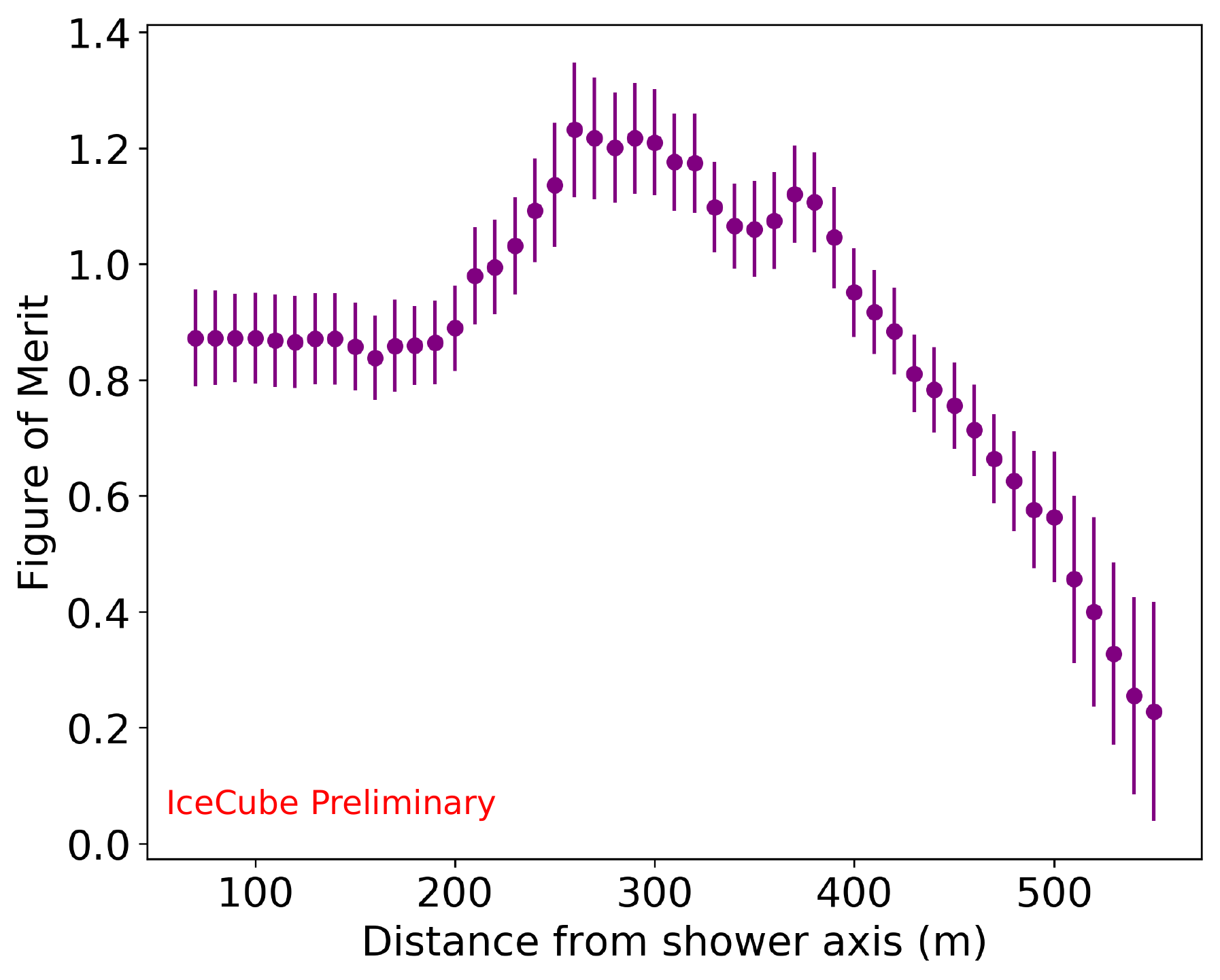}
\caption{Left: An example of the FOM calculation for showers with $\theta\,=\,55^\circ-60^\circ$ and log$_{10}$(S$_{125}$/VEM)$\,=\,0.7-1.0$. A Gaussian distribution is fit to both iron and proton distributions of $ \xi_{\mu-\mathrm{like}}$ for a given distance bin (here 260 m). Right: The obtained FOM vs distance for showers with $\theta\,=\,50^\circ-55^\circ$ and log$_{10}$(S$_{125}$/VEM)$\,=\,1.0-1.3$.  The reference distance is chosen from this.}\label{fig4}
\end{figure}

An example of the FOM calculation is given in Figure \ref{fig4} (left). The figure shows the distributions of $ \xi_{\mu-\mathrm{like}}$ at a distance of 260 m for proton and iron showers with $\theta\,=\,55^\circ-60^\circ$ and with log$_{10}$(S$_{125}$/VEM)$\,=\,0.7-1.0$. The resulting value of the FOM for this distribution is 1.08 $\pm$ 0.08. Once the FOM is calculated for each distance bin, they can be compared as shown in Figure \ref{fig4} (right). The figure shows the FOM values at different distances for $\theta\,=\,50^\circ-55^\circ$ and log$_{10}$(S$_{125}$/VEM)$\,=\,1.0-1.3$. Based on this, the distance of 260 m is chosen as the reference distance for this bin. This procedure is repeated for all zenith bins and all energy bins used in the analysis. 

\begin{figure}[h]
\centering
\includegraphics[width=.7\textwidth]{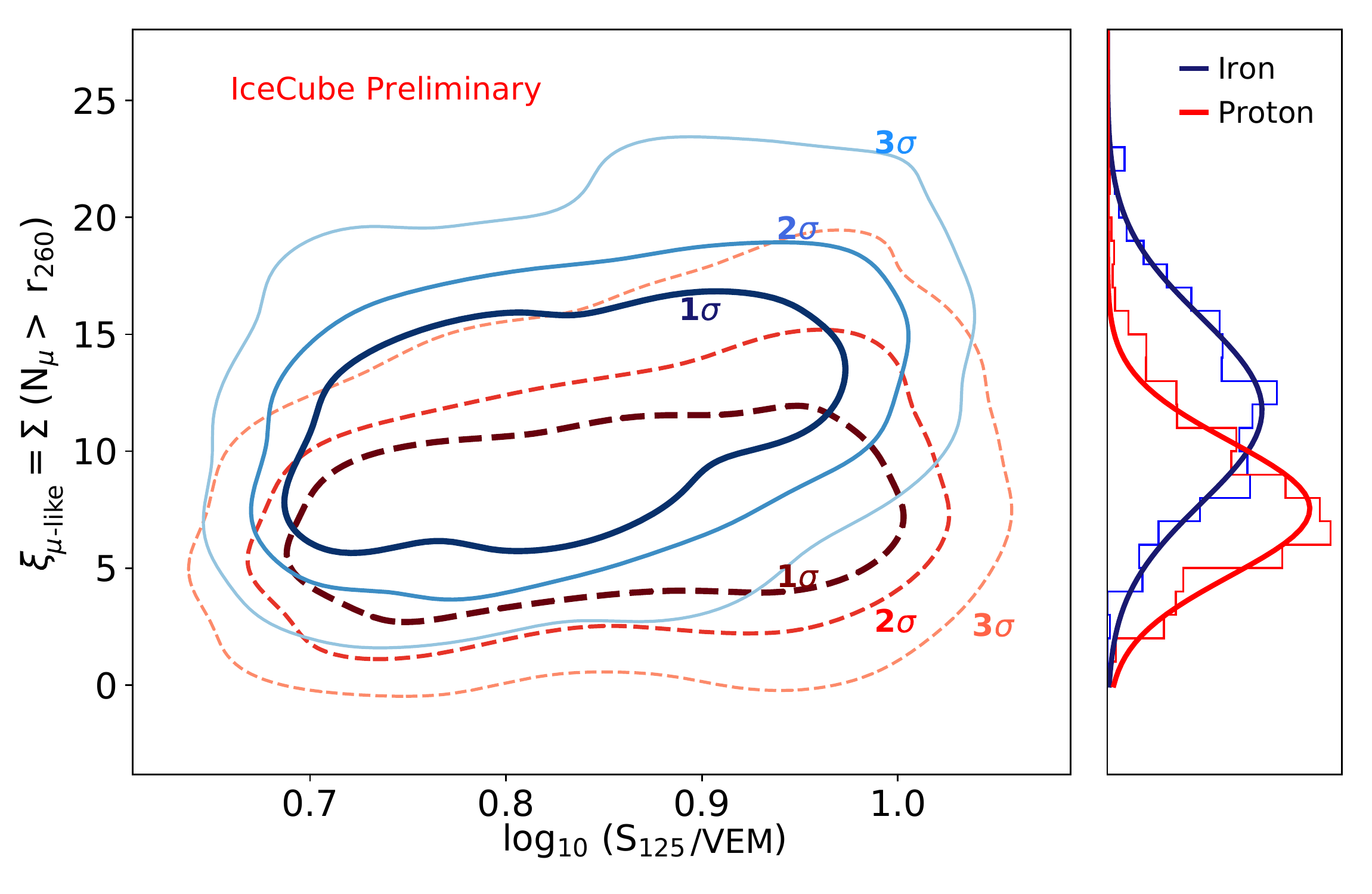}
\caption{Probability-density distributions for proton (red dashed) and iron (blue solid) showers with $\theta\,=\,50^\circ-55^\circ$ and log$_{10}$(S$_{125}$/VEM)$\,=\,0.7-1.0$ for the reference distance of 260 m. The right panel shows the combined distribution of $ \xi_{\mu-\mathrm{like}}$.}\label{fig5}
\end{figure}

Upon the determination of the reference distance, the probability-density  distribution for $ \xi_{\mu-\mathrm{like}}$ can be drawn for both proton and iron showers. This is done for each zenith and energy bin considered in the analysis. An example of such a distribution for showers with $\theta\,=\,50^\circ-55^\circ$ and log$_{10}$(S$_{125}$/VEM)$\,=\,0.7-1.0$ is shown in Figure \ref{fig5}. The figure shows the two-dimensional probability-density distribution of $ \xi_{\mu-\mathrm{like}}$ at the reference distance of 260 m with respect to log$_{10}$(S$_{125}$/VEM). The solid-blue curves in the figure represent the $1\sigma$, $2\sigma$ and $3\sigma$ contours for the iron showers while the red-dashed curves represent the same for the proton showers. The right panel in the figure also shows the combined $ \xi_{\mu-\mathrm{like}}$ distributions of proton and iron showers within the entire bin. A separation between the iron and proton showers is visible, with some amount of overlap. This signifies that in the regions with less overlap, we can identify the showers more confidently, while in the overlap regions the shower primary becomes more ambiguous.

The nature of the iron and proton probability-density distributions is utilised to evaluate the light or heavy nature of each observed air-shower event. For this, the marginal cumulative distribution (summation along $ \xi_{\mu-\mathrm{like}}$-axis) is determined along each slice of log$_{10}$(S$_{125}$). The summation is performed along the positive $y$-axis for iron and along the negative $y$-axis for the proton distribution. This results in an iron-marginal cumulative distribution (Fe-CD) value and proton-marginal cumulative distribution (H-CD) value for each point in this space. With this, we can assign a Fe-CD and H-CD value to each event with a given pair of log$_{10}$(S$_{125}$) and $ \xi_{\mu-\mathrm{like}}$. Any event with a high H-CD value and low Fe-CD value would be identified as light (or proton-like) and vise-versa for heavy (or iron-like) events.

\section{Testing the procedure}
In order to verify the performance of the analysis procedure, we conduct a test using simulations. The simulated air showers are passed through the analysis to see the fraction of proton/iron events correctly identified as light/heavy.

\begin{figure}[h]

\includegraphics[width=.5\textwidth]{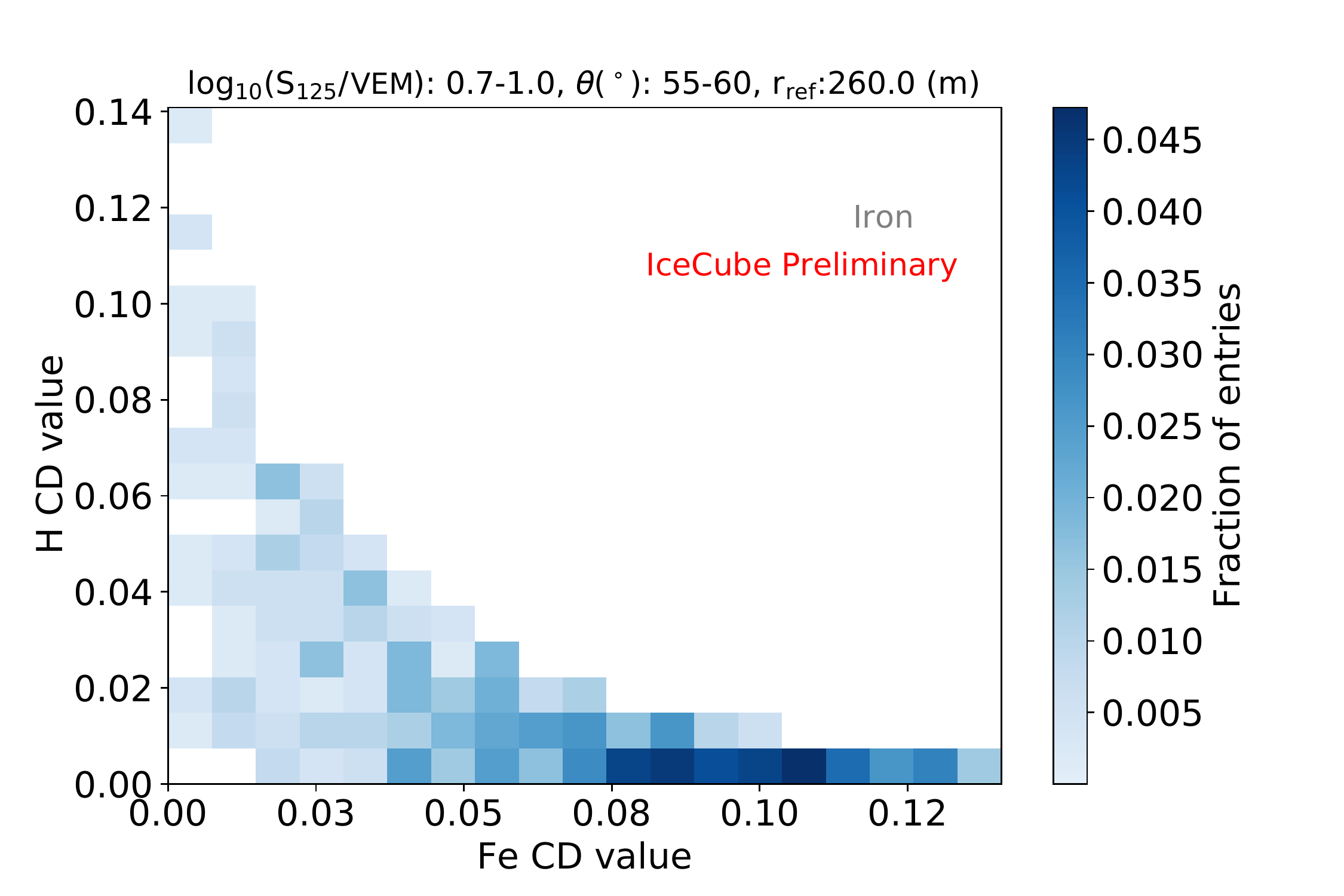}
\includegraphics[width=.49\textwidth]{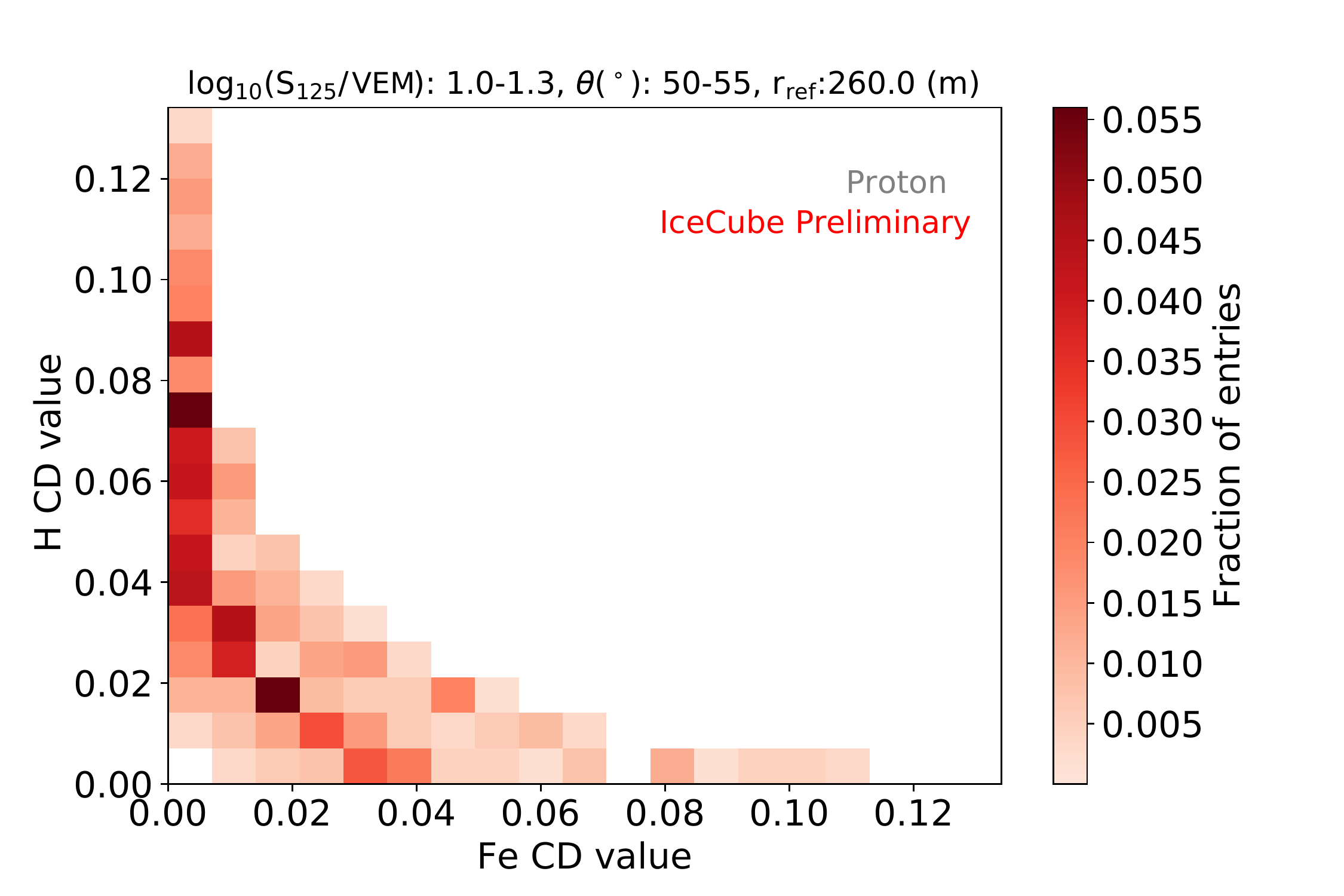}
\caption{Verification of the mass-classification procedure using simulations. The H-CD value and Fe-CD value is created for each event and compared. Left: Iron showers (log$_{10}$(S$_{125}$/VEM) = 0.7-1.0 and $\theta\,=\,55-60^\circ$), all events falling in the bins with low H-CD values are correctly identified as heavy. 45.4\% events fall in the 1st row (correctly identified) and 2.7\% events fall in the 1st column (misidentified as light). Right: Proton showers (log$_{10}$(S$_{125}$/VEM) = 1.0-1.3 and $\theta$ = 50-55$^\circ$), with events falling in the bins with low Fe-CD values identified as light. 46.2\% events fall in the 1st row (correctly identified) and 12.9\% events fall in the 1st column (misidentified as heavy)}\label{fig6}
\end{figure}

Figure \ref{fig6} shows the verification done for iron (left) and proton (right) simulations, on an event-by-event basis. Each simulated event is assigned an H-CD and Fe-CD value, and a histogram of these assigned values is shown in the figure. The colour scale depicts the fraction of events entering each bin in the histogram. Events with low Fe-CD values  are considered as light and events with low H-CD values are considered as heavy. 

We can sum up the fractions in the bottom-first row of bins on the x-axis (with low H-CD value) to determine the percentage of entries identified as heavy. The same can be done for each set of rows. Similarly, the columns can be added up to see the percentage of entries in each column, the leftmost column (with low Fe-CD) being the events most confidently identified as light. 

Table \ref{tab:tab1} shows the percentage of events that enter the bottom-first row (with low H-CD value) and the bottom-second row, that are identified as heavy for showers with $\theta\,=50-55^\circ$ and log$_{10}$(S$_{125}$/VEM) = 0.7-1.0 (middle $\theta$ and energy bins). The table also shows the percentage of events in the leftmost column (with low Fe-CD value) and the second column from the left, that are identified as light. The same calculation can be performed for the events shown in Figure \ref{fig6} also.

It is clear that $40-50\%$ of iron events are correctly identified as heavy (with better confidence)
and $30-40\%$ of proton events are correctly identified as light. A smaller percentage of proton/iron events are incorrectly identified as heavy/light.
One caveat to this is that the same events used for drawing the probability-density distributions have been used to test the procedure, due to the lack of sufficient simulations.

\begin{figure}[h]

\includegraphics[width=.5\textwidth]{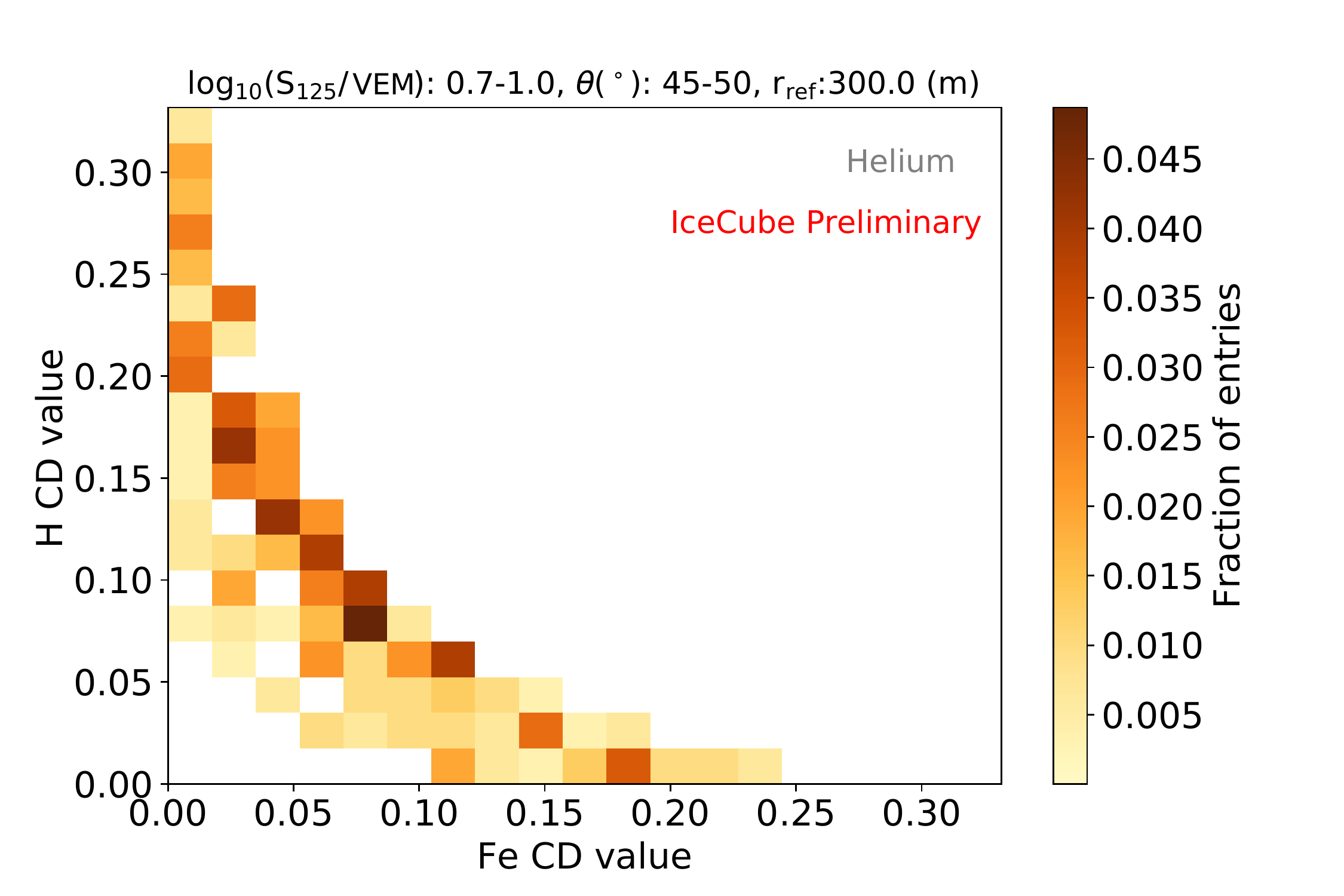}
\includegraphics[width=.49\textwidth]{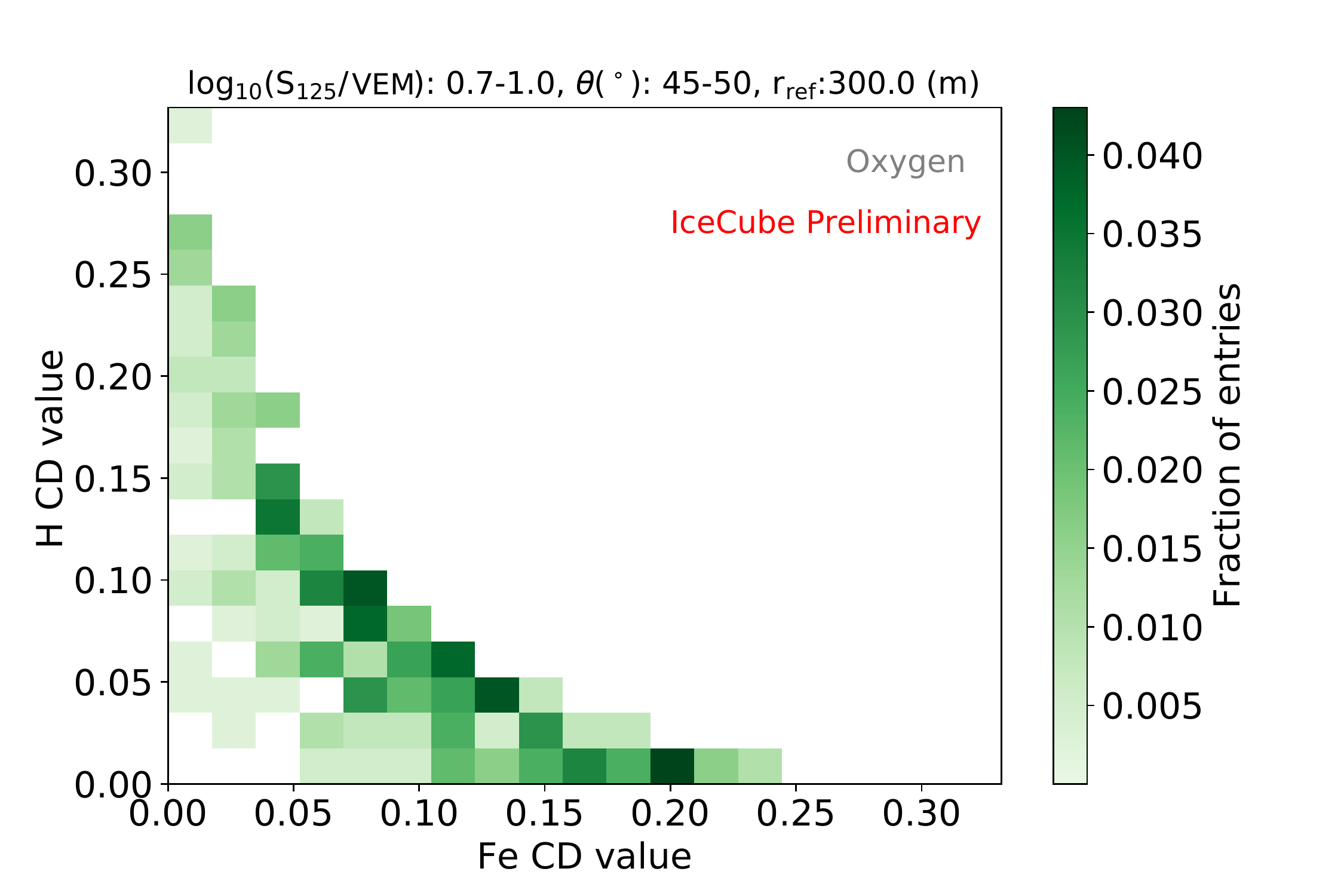}
\caption{The verification done for helium showers (left) and oxygen showers (right) with, log$_{10}\mathrm{S_{125}/VEM)}\,=\,0.7-1.0$ and $\theta\,=\,45-50^\circ$, using the H-CD and Fe-CD values obtained from the probability-density distributions obtained from proton and iron primaries. 10.0\% He events fall in the 1st row (heavy) and 17.2\% events fall in the 1st column (light). 20.4\% He events fall in the 1st row (heavy) and 7.8\% events fall in the 1st column (light).}\label{fig7}
\end{figure}

The same procedure can be repeated for the available helium and oxygen simulations to test the classifier.
This is shown in Figure \ref{fig7} for helium and oxygen events with $\theta\,=\,45^\circ-50^\circ$ and log$_{10}$(S$_{125}$/VEM)$\,=\,0.7-1.0$. 
The helium and oxygen events are not seen to clump as closely to the light/heavy composition bins as the proton/iron events. 
They show mixed nature, as expected. This shows that the classification is indeed composition sensitive since helium and oxygen should fall as intermediaries.  
The percentage of helium and oxygen events that fall in the light-identified columns and heavy-identified rows for air showers falling in the middle $\theta$ and S$_{125}$ bins are shown in \mbox{Table \ref{tab:tab1}}. 

\begin{table}[]
\centering
\begin{tabular}{|c|c|c|c|c|}
\hline
\begin{tabular}[c]{@{}c@{}}Element\end{tabular} & \begin{tabular}[c]{@{}c@{}}Events in \\ 1st column\\ (light)\end{tabular} & \begin{tabular}[c]{@{}c@{}}Events in \\ 2nd column\\ (light)\end{tabular} & \begin{tabular}[c]{@{}c@{}}Events in \\ 1st row\\ (heavy)\end{tabular} & \begin{tabular}[c]{@{}c@{}}Events in \\ 2nd row\\ (heavy)\end{tabular} \\ \hline
\begin{tabular}[c]{@{}c@{}}True H\\ \end{tabular}                         & 26.7\%                                                                                 & 14.6\%                                                                                 & 6.2\%                                                                              & 7.1\%                                                                              \\ \hline
\begin{tabular}[c]{@{}c@{}}True Fe\\ \end{tabular}                        & 3.3\%                                                                                  & 4.4\%                                                                                  & 40.2\%                                                                              & 17.1\%                                                                              \\ \hline
\begin{tabular}[c]{@{}c@{}}True He\\ \end{tabular}                        & 14.7\%                                                                                 & 10.6\%                                                                                 & 16.1\%                                                                              & 11.6\%                                                                               \\ \hline
\begin{tabular}[c]{@{}c@{}}True O\\ \end{tabular}                         & 7.1\%                                                                                  & 6.2\%                                                                                  & 21.0\%                                                                              & 13.6\%                                                                              \\ \hline
\end{tabular}
\caption{Testing the classifier for various elements for events that fall in the middle $\theta$ and energy proxy bins $\theta\,=\,50-55^\circ$ and log$_{10}$(S$_{125}$/VEM) = 1.0-1.3.}\label{tab:tab1}
\end{table}

\section{Summary and Outlook}\label{sec:refs}
We have explored a new method for determining the composition of inclined air showers on an event-by-event basis. This utlilizes the discrimination of muon-like signals in the air shower and using it to draw probability-density distributions for both proton and iron showers. Based on this, we classify events as light or heavy by assigning them marginal cumulative distribution values and marginal inverse cumulative distribution values. A verification of the procedure reveals that the method is effective.
Further tests on other hadronic interaction models will be needed to determine the model-dependent performance of the analysis. With that, we can use the analysis method to determine the mass composition of inclined air showers measured with IceTop, and thereby provide an independent check to the composition measurement of quasi-vertical showers from the same detector array.

\bibliographystyle{ICRC}
\bibliography{references}



\clearpage
\section*{Full Author List: IceCube Collaboration}



\scriptsize
\noindent
R. Abbasi$^{17}$,
M. Ackermann$^{59}$,
J. Adams$^{18}$,
J. A. Aguilar$^{12}$,
M. Ahlers$^{22}$,
M. Ahrens$^{50}$,
C. Alispach$^{28}$,
A. A. Alves Jr.$^{31}$,
N. M. Amin$^{42}$,
R. An$^{14}$
% ...
\\

\noindent
% ...
$^{12}$ Universit{\'e} Libre de Bruxelles, Science Faculty CP230, B-1050 Brussels, Belgium \\
% ...
$^{14}$ Department of Physics and Laboratory for Particle Physics and Cosmology, Harvard University, Cambridge, MA 02138, USA \\
% ...
$^{17}$ Department of Physics, Loyola University Chicago, Chicago, IL 60660, USA \\
$^{18}$ Dept. of Physics and Astronomy, University of Canterbury, Private Bag 4800, Christchurch, New Zealand \\
% ...
$^{22}$ Niels Bohr Institute, University of Copenhagen, DK-2100 Copenhagen, Denmark \\
% ...
$^{28}$ D{\'e}partement de physique nucl{\'e}aire et corpusculaire, Universit{\'e} de Gen{\`e}ve, CH-1211 Gen{\`e}ve, Switzerland \\
% ...
$^{31}$ Karlsruhe Institute of Technology, Institute for Astroparticle Physics, D-76021 Karlsruhe, Germany  \\
% ...
$^{42}$ Bartol Research Institute and Dept. of Physics and Astronomy, University of Delaware, Newark, DE 19716, USA \\
% ...
$^{50}$ Oskar Klein Centre and Dept. of Physics, Stockholm University, SE-10691 Stockholm, Sweden \\
% ...
$^{59}$ DESY, D-15738 Zeuthen, Germany \\
% ...

\subsection*{Acknowledgements}

\noindent
USA {\textendash} U.S. National Science Foundation-Office of Polar Programs,
U.S. National Science Foundation-Physics Division,
U.S. National Science Foundation-EPSCoR,
Wisconsin Alumni Research Foundation,
Center for High Throughput Computing (CHTC) at the University of Wisconsin{\textendash}Madison,
Open Science Grid (OSG),
Extreme Science and Engineering Discovery Environment (XSEDE),
Frontera computing project at the Texas Advanced Computing Center,
U.S. Department of Energy-National Energy Research Scientific Computing Center,
Particle astrophysics research computing center at the University of Maryland,
Institute for Cyber-Enabled Research at Michigan State University,
and Astroparticle physics computational facility at Marquette University;
Belgium {\textendash} Funds for Scientific Research (FRS-FNRS and FWO),
FWO Odysseus and Big Science programmes,
and Belgian Federal Science Policy Office (Belspo);
Germany {\textendash} Bundesministerium f{\"u}r Bildung und Forschung (BMBF),
Deutsche Forschungsgemeinschaft (DFG),
Helmholtz Alliance for Astroparticle Physics (HAP),
Initiative and Networking Fund of the Helmholtz Association,
Deutsches Elektronen Synchrotron (DESY),
and High Performance Computing cluster of the RWTH Aachen;
Sweden {\textendash} Swedish Research Council,
Swedish Polar Research Secretariat,
Swedish National Infrastructure for Computing (SNIC),
and Knut and Alice Wallenberg Foundation;
Australia {\textendash} Australian Research Council;
Canada {\textendash} Natural Sciences and Engineering Research Council of Canada,
Calcul Qu{\'e}bec, Compute Ontario, Canada Foundation for Innovation, WestGrid, and Compute Canada;
Denmark {\textendash} Villum Fonden and Carlsberg Foundation;
New Zealand {\textendash} Marsden Fund;
Japan {\textendash} Japan Society for Promotion of Science (JSPS)
and Institute for Global Prominent Research (IGPR) of Chiba University;
Korea {\textendash} National Research Foundation of Korea (NRF);
Switzerland {\textendash} Swiss National Science Foundation (SNSF);
United Kingdom {\textendash} Department of Physics, University of Oxford.
\end{document}